\documentclass[12pt,a4paper,english,superscriptaddress,aps,nofootinbib]{revtex4}

\usepackage{amsmath,amssymb,amsfonts,bm}
\usepackage{slashed}
\usepackage{graphicx}
\usepackage{hyperref}
\usepackage{subfigure}
\usepackage{tikz}
\usepackage{tikz-feynman}
\usepackage{mathtools}

\begin{document}

\title{From Dirac surface response to Casimir forces in Bumblebee-CFJ electrodynamics}

\author{F. C. E. Lima}
\affiliation{Centro de Matemática, Computação e Cognição (CMCC), Universidade Federal do ABC (UFABC), Av. dos Estados 5001, CEP 09210-580, Santo André, São Paulo, Brazil}

\begin{abstract}
\noindent \textbf{Abstract:} We examine the Casimir interaction between planar massive Dirac materials coupled to a Lorentz-violating Bumblebee-CFJ vector sector. The standard renormalized one-loop polarization tensor of the Dirac quasiparticles is used as the nonlocal surface response in the Lifshitz scattering formalism. For the normally oriented Lorentz-violating background, the surface response remains parity-even. Hence, the repulsive regimes do not originate from a Hall-like conductivity. Instead, they arise from the CFJ-induced mixing of TE and TM bulk modes, combined with the spatially dispersive response of the Dirac sheets. Furthermore, we show that the Dirac mass suppresses the interaction and changes its large-distance scaling, whereas the CFJ parameter drives transitions between attractive and repulsive regimes.

\end{abstract}

\maketitle

%%%%%%%%%%%%%%%%%%%%%%%%%%%%%%%%%%%%%%%%%%%%%%%%%%%%%%%%%%%%%%%%%%%%%%%%%%%%%%%%%%%%%%%%%%%%%%%%%%%

\section{Introduction}
\label{sec:introduction}

The Casimir interaction probes both the propagation of quantum fluctuations and the response of the bodies that scatter them. Starting with the force between perfectly conducting plates \cite{Casimir1948}, Lifshitz theory extended the description to dispersive materials \cite{Lifshitz1956,Dzyaloshinskii1961}, establishing the connection between vacuum forces and electromagnetic response functions \cite{Bordag2001,Bordag2009,Dalvit2011}. In the scattering formulation, propagation through the intervening region and reflection by each body enter separately in an interaction determinant \cite{Kenneth2008}. This separation provides a natural background for investigating how microscopic material properties and modified field dynamics jointly determine fluctuation-induced forces \cite{Buhmann2012,Woods2016}.

Planar Dirac materials allow microscopic derivation of the sheets' response. In this framework, the low-energy quasiparticles are described by an effective Dirac Hamiltonian with a Fermi velocity $v_F$, which is distinct from the standard electromagnetic velocity \cite{Geim2007,CastroNeto2009,Katsnelson2012}. Thereby, optical conductivity, screening, and collective excitations depend on the electronic spectrum and its occupation \cite{GusyninSharapovCarbotte2007,Wunsch2006,Hwang2007}, with the mass gap introducing an interband threshold and suppressing the low-energy response \cite{Pyatkovskiy2009}. The polarization tensor incorporates these effects as a function of frequency and in-plane momentum, providing the response functions needed to construct reflection coefficients \cite{Fialkovsky2011,Bordag2015Tensor}. Thus, it preserves its spatial dispersion to distinguish the response sampled by a Casimir interaction from the homogeneous optical limit.

The resulting Casimir interaction has been studied for graphene opposite a conductor \cite{Bordag2015}, graphene coupled to metamaterial structures \cite{Drosdoff2012}, and doped or thermally excited sheets \cite{Fialkovsky2011,BordagFialkovskyVassilevich2016,Klimchitskaya2014}. These studies establish that the gap, carrier density, and temperature can substantially change the interaction without modifying Maxwell propagation. In related graphene-family materials, different electronic phases produce distinct distance dependences and allow quantized or repulsive interactions \cite{RodriguezLopez2016}. Measurements involving graphene-coated substrates and the observation of an enhanced thermal contribution provide experimental benchmarks for the material-response description \cite{Banishev2013,Klimchitskaya2021}. They motivate a microscopic treatment of the sheets but do not, by themselves, establish a Lorentz-violating propagation sector.

Parity-odd surface response provides an important reference point for the present problem. In planar Dirac systems, Chern-Simons and Hall contributions may arise from parity-breaking fermionic masses and have been explored as mechanisms for modifying and, in some configurations, reversing Casimir forces \cite{Niemi1983,Redlich1984,Deser1982,Semenoff1984,Haldane1988,Kane2005,Hasan2010,Qi2011,Wilczek1987,Qi2008,Essin2009,Grushin2011,Tse2012,Fialkovsky2018}. However, such effects originate from the response of the material surface itself. In the configuration considered here, the parity-odd contributions of the two valleys cancel, leaving a parity-even surface polarization tensor. Any polarization mixing or repulsive behavior must therefore originate from the modified bulk propagation, providing a mechanism physically distinct from Hall-induced Casimir repulsion.

In our the system, the planar Dirac sheets do not generate a net Hall response, since the parity-odd contributions from the two valleys cancel in ordinary graphene. Hence, the surface response entering the Lifshitz problem is parity even and coincides with the standard massive Dirac polarization tensor, with the electromagnetic coupling replaced by the effective Bumblebee coupling. Therefore, any repulsive regime found in the force must have a different origin. As we show below, it arises from the CFJ-induced mixing of TE and TM modes in the bulk propagation, together with the nonlocal spatially dispersive response of the Dirac sheets.

Lorentz-violating effective field theories provide a systematic setting for modifying the propagation sector. The Standard-Model Extension (SME) organizes Lorentz- and CPT-violating operators in matter and gauge theories \cite{Colladay1997,Kostelecky1999,Kostelecky2004}. In electrodynamics, the Carroll-Field-Jackiw (CFJ) term couples a fixed four-vector to the gauge field and its dual field strength, modifying dispersion and polarization \cite{CarrollFieldJackiw1990}. Several studies analyzed the possible radiative generation of the CFJ term and its sensitivity to the regularization prescription \cite{Jackiw1999,PerezVictoria2001}. Consistency depends on the background and the effective-theory regime, as illustrated by analyses of stability and causality in Refs. \cite{Lehnert2004,Adam2002}. Furthermore, birefringence and other photon-sector observables also impose strong phenomenological constraints \cite{KosteleckyMewes2002,Kostelecky2011}. Accordingly, a CFJ coefficient entering a material-based model must be distinguished from an experimentally established property of the vacuum.

Bumblebee models provide a dynamical realization of a preferred direction through a vector Vacuum Expectation Value (VEV) \cite{Kostelecky2004b,BluhmKostelecky2005,Bluhm2008}. Their fluctuation content depends on the kinetic term, the symmetry-breaking potential, and the constraints imposed on the admissible sector \cite{BluhmGagne2008,Seifert2010}. Moreover, quantum corrections require distinguishing the full vector theory from a purely transverse photon-like sector \cite{Maluf2015}. The relation to electrodynamics can be established under appropriate conditions, including the restricted sector associated with a nonlinear gauge \cite{Escobar2017}. Thus, identifying a response to Bumblebee fluctuations with an ordinary electromagnetic conductivity is an additional physical assumption, not a consequence of the planar Dirac description alone.

Modified bulk propagation is already known to affect Casimir forces. Calculations in Maxwell-Chern-Simons electrodynamics demonstrate how Lorentz-violating dispersion modifies the Casimir interaction \cite{KharlanovZhukovsky2010}. Notably, a spatial axion gradient can produce attractive and repulsive regimes between parallel plates \cite{cas-fukushima}. Recent work also studies axion-induced pressure in graphene cavities driven by axion dark matter in an external magnetic field \cite{Alachkar2026}. Naturally, that driven setting differs from the equilibrium vacuum interaction adopted in our project. Indeed, these results motivate the more specific question addressed here, i.e., how do the finite reflectivity, mass gap, and full spatial dispersion of Dirac sheets modify the force mediated by a homogeneous CFJ bulk? This question will be addressed throughout the present analysis.

The paper is organized as follows. Section \ref{sectionII} introduces the Bumblebee-CFJ framework and its coupling to planar Dirac matter. Section \ref{sectionIII} presents the fermionic propagator and interaction vertices. Section \ref{sectionIV} discusses the one-loop polarization tensor and the homogeneous response and its conductivity interpretation. Section \ref{sectionVI} constructs the bulk modes and reflection matrices, derives the Casimir energy and pressure, and discusses the limiting cases and numerical results. Section \ref{sectionVII} contains our conclusions.

\section{Effective Bumblebee-CFJ Framework for Planar Dirac Matter}\label{sectionII}

To describe the interplay between a planar Dirac material and a Lorentz-violating vector sector, we adopt the action, viz.,
\begin{align}
S=S_B+S_\psi ,
\label{eq:S-total}
\end{align}
where
\begin{align}
S_B=\int d^4x\,\left[-\frac14 B_{\mu\nu}B^{\mu\nu}+\frac12(\kappa_{AF})_\mu\epsilon^{\mu\nu\rho\sigma}B_\nu B_{\rho\sigma}-V\!\left(B_\mu B^\mu+ b^2\right)\right],
\label{eq:SB}
\end{align}
and
\begin{align}
S_\psi=\int\, d^3x\,\bar\psi\,\slashed{D}_B\,\psi ,
\label{eq:Spsi}
\end{align}
Here, we have a minimal effective framework that consistently incorporates both the bulk dynamics of the mediating field and the quantum response of the quasiparticles confined to the sheet. The bosonic contribution $S_B$ is formulated in $(3+1)$ dimensions and combines the standard Maxwell-like kinetic term with a Bumblebee potential, responsible for selecting a nonvanishing vector VEV and inducing spontaneous Lorentz-symmetry breaking, together with a CPT-odd CFJ term that parametrizes an additional explicit Lorentz-violating contribution. Meanwhile, $S_\psi$ describes the $(2+1)$-dimensional massive Dirac quasiparticles localized at $x^{3}=0$, including the characteristic Fermi velocity $v_F$ and their coupling to the tangential components of the Bumblebee field. This construction is particularly suitable for the present problem, since it preserves the established Dirac description of the material response while allowing the propagation sector mediating the Casimir interaction to acquire the nontrivial vacuum structure.

Since the Dirac quasiparticles are confined to the $(2+1)$-dimensional plane, only the temporal and tangential components of the Bumblebee field, i.e., $B_{0}$, $B_{1}$, and $B_{2}$, couple directly to the fermionic degrees of freedom. Therefore, the Dirac operator describing the fermionic sector takes the form\footnote{In this work, we assume that the graphene sheet is located at $x^3=0$.  Furthermore, we consider $g_{\mu\nu}=\mathrm{diag}\,(+,-,-,-)$, $\epsilon^{0123}=+1$, and $B_{\mu\nu}=\partial_\mu B_\nu-\partial_\nu B_\mu$. Consequently, for the $(2+1)$-dimensional Dirac algebra, one has $g_{mn}=\mathrm{diag}(+,-,-)$ and $m,n=0,1,2$.}
\begin{align}
\slashed{D}_B=\gamma^0(i\partial_0-gB_0)+v_F\gamma^1(i\partial_1-gB_1)+v_F\gamma^2(i\partial_2-gB_2)-m .
\label{eq:Dirac-operator}
\end{align}

For the Lorentz-violating vacuum, we choose a constant space-like background oriented along the direction normal to the Dirac sheet\footnote{Therefore, one has $b^\mu=(0,0,0,-b)$, which leads us to $b_\mu\,b^\mu=-b^2$.}, i.e., 
\begin{align}
b_\mu=(0,0,0,b) \qquad \mathrm{and} \qquad (\kappa_{AF})_\mu=(0,0,0,\kappa).
\label{eq:backgrounds}
\end{align}

Upon adopting the metric convention $g_{\mu\nu}=\mathrm{diag}(+,-,-,-)$, this background satisfies $b_\mu b^\mu=-b^2$. Accordingly, the Bumblebee potential is taken to depend on the invariant combination $X \equiv B_\mu B^\mu+b^2$, such that its minimum defines the Lorentz-violating vacuum, $\left. X \right|_{B^\mu=b^\mu}=0$. Upon assuming that the potential is stationary and locally stable at this minimum, one has $V'(0)=0$ and $V''(0)>0$. The nonvanishing VEV $\langle B^\mu\rangle=b^\mu$ selects a preferred space-like direction and spontaneously breaks Lorentz symmetry.

To accomplish our purpose, we will study the Dirac description and the polarization tensor calculation following the strategy of Ref. \cite{Fialkovsky2011}.  The spontaneous Lorentz breaking associated with a vector VEV is characteristic of Bumblebee models \cite{BluhmKostelecky2005}, while the CPT-odd term in Eq. \eqref{eq:SB} has the Carroll-Field-Jackiw structure \cite{CarrollFieldJackiw1990}.

%%%%%%%%%%%%%%%%%%%%%%%%%%%%%%%%%%%%%%%%%%%%%%%%%%%%%%%%%%%%%%%%%%%%%%%%%%%%%%%%%%%%%%%%%%%%%%

\section{On the Fermionic sector}
\label{sectionIII}

Once the Bumblebee field propagates in the full $(3+1)$-dimensional spacetime, the Dirac quasiparticles are restricted to the plane $x^3=0$. Therefore, one can couple directly only to the pullback of the vector field onto the sheet, namely its components $B_0$, $B_1$, and $B_2$. Naturally, this observation clarifies the role played by the Lorentz-violating vacuum in the matter sector. Since the VEV has been chosen along the direction normal to the sheet, the background $b^\mu$ does not act as a direct Lorentz-violating deformation of the planar Dirac Hamiltonian. Instead, its effects on the Casimir interaction arise through the modified propagation of the Bumblebee field in the bulk. Thus, the fermionic sector determines the quantum response of the material surface, whereas the Bumblebee-CFJ sector determines how the corresponding fluctuations propagate between the surfaces.

To make this structure explicit, we decompose the Bumblebee field as $B_\mu=b_\mu+\delta B_\mu$, where $b_\mu$ denotes the constant Lorentz-violating vacuum and $\delta B_\mu$ represents the dynamical fluctuation around it. Substituting this decomposition into the Dirac operator introduced in Eq. \eqref{eq:Dirac-operator}, one obtains
\begin{align}
\slashed{D}_B=&\gamma^0\left(i\partial_0-gb_0-g\delta B_0\right)+v_F\gamma^1\left(i\partial_1-gb_1-g\delta B_1\right)+v_F\gamma^2\left(i\partial_2-gb_2-g\delta B_2\right)-m.
\label{34}
\end{align}
Adopting the space-like vacuum configuration, i.e., $b^\mu=(0,0,0,-b)$, the components tangent to the Dirac sheet vanish identically, i.e., $b_0=b_1=b_2=0$. Consequently, the fermionic operator boils down to 
\begin{align}
\slashed{D}_B=\gamma^0\left(i\partial_0-g\delta B_0\right)+v_F\gamma^1\left(i\partial_1-g\delta B_1\right)+v_F\gamma^2\left(i\partial_2-g\delta B_2\right)-m.
\label{37}
\end{align}
Note that Eq. \eqref{37} has an important physical interpretation. The preferred direction selected by spontaneous Lorentz-symmetry breaking is orthogonal to the plane in which the quasiparticles propagate. Thus, a nonvanishing temporal component $b_0$ would directly shift the temporal part of the Dirac operator, whereas nonzero tangential components $b_1$ or $b_2$ would modify the in-plane momentum structure. The absence of such terms in Eq. \eqref{37} is a consequence of the normal orientation of the Bumblebee vacuum.

\subsection{Free fermion propagator}

Firstly, let us determine the propagator describing the unperturbed quasiparticles. For this purpose, one notes that the external Bumblebee fluctuations are temporarily set to zero, i.e., $\delta B_0=\delta B_1=\delta B_2=0$. Thereby, the resulting operator defines the free fermionic theory around which the effective action will subsequently be expanded.

Upon using the momentum convention employed throughout the polarization calculation\footnote{It is worth emphasizing that the convention adopted here is $\psi(x)\sim \mathrm{e}^{-iq\cdot x}$, where $q$ is the internal fermion momentum.}, the inverse fermion propagator is
\begin{align}
S^{-1}(q_0,q_1,q_2)=q_0\gamma^0+v_Fq_1\gamma^1+v_Fq_2\gamma^2-m .
\label{38}
\end{align}
Note that  the spatial components are multiplied by $v_F$, reflecting the fact that the quasiparticles propagate with the effective Fermi velocity rather than with the relativistic velocity $c=1$. Therefore, although the Dirac algebra remains relativistic in form, the quasiparticle dispersion is anisotropically rescaled between the temporal and spatial sectors.

To invert Eq. \eqref{38}, we multiply the Dirac operator by the corresponding operator with the opposite sign of the mass, i.e., $q_0\gamma^0+v_Fq_1\gamma^1+v_Fq_2\gamma^2+m$. Considering the relevant three-dimensional Clifford algebra, i.e., $\{\gamma^a,\gamma^b\}=2g^{ab}$ and $g^{ab}=\mathrm{diag}(+,-,-)$ with $a,b=0,1,2$. 

Naturally, this approach leads us to
%\begin{align}\nonumber
%\Big[q_0\gamma^0+v_Fq_1\gamma^1+v_Fq_2\gamma^2-m\Big]&\Big[q_0\gamma^0+v_Fq_1\gamma^1+v_Fq_2\gamma^2+m\Big]=\\ \nonumber
%=&q_0^2(\gamma^0)^2+v_F^2q_1^2(\gamma^1)^2+v_F^2q_2^2(\gamma^2)^2-m^2+v_Fq_0q_1\left(\gamma^0\gamma^1+\gamma^1\gamma^0\right)\\ \nonumber
%+&v_Fq_0q_2\left(\gamma^0\gamma^2+\gamma^2\gamma^0\right)+v_F^2q_1q_2\left(\gamma^1\gamma^2+\gamma^2\gamma^1\right).
%=&q_0^2-v_F^2q_1^2-v_F^2q_2^2-m^2.
%\label{41}
%\end{align}
%Therefore, taking into account that $(\gamma^0)^2=+1$ and $(\gamma^1)^2=(\gamma^2)^2=-1$, together with $\{\gamma^0,\gamma^1\}=\{\gamma^0,\gamma^2\}=\{\gamma^1,\gamma^2\}=0$\footnote{Thus, the mixed terms cancel by virtue of the Clifford algebra, yielding a scalar denominator.}, one obtains 
%\begin{align}
%    \Big[q_0\gamma^0+v_Fq_1\gamma^1+v_Fq_2\gamma^2-m\Big]\Big[q_0\gamma^0+v_Fq_1\gamma^1+v_Fq_2\gamma^2+m\Big]=q_0^2-v_F^2q_1^2-v_F^2q_2^2-m^2.
%\end{align}
%Therefore, within this framework, one obtains\footnote{Furthermore, the infinitesimal term $i0^+$ in Eq. \eqref{42} specifies the causal prescription of the Minkowski-space propagator and does not represent an independent physical parameter.}
\begin{align}
S(q_0,q_1,q_2)=\frac{q_0\gamma^0+v_Fq_1\gamma^1+v_Fq_2\gamma^2+m}{q_0^2-v_F^2q_1^2-v_F^2q_2^2-m^2+i0^+}.
\label{42}
\end{align}
The pole structure of Eq. \eqref{42} provides a direct consistency check of the fermionic propagator, since its poles must reproduce the quasiparticle dispersion implied by the free Dirac operator. Therefore, by adopt the denominator of Eq. \eqref{42} equal to zero gives
\begin{align}
E_{\mathbf{q}}=\sqrt{m^2+v_F^2(q_1^2+q_2^2)},
\end{align}
and the poles satisfy $q_0=\pm E_{\mathbf q}$. Accordingly, $m$ controls the Dirac gap, whereas $v_F$ fixes the slope of the in-plane dispersion.

\subsection{Fermion-Bumblebee interaction vertices}

Now, let us isolate the part of Eq. \eqref{37} that couples the fermionic current to the Bumblebee fluctuations. Subtracting the free Dirac operator from Eq. \eqref{37}, the interaction Lagrangian reads
\begin{align}
\mathcal{L}_{\mathrm{int}}=-g\bar{\psi}\gamma^0\psi\,\delta B_0-gv_F\bar{\psi}\gamma^1\psi\,\delta B_1-gv_F\bar{\psi}\gamma^2\psi\,\delta B_2.
\label{43}
\end{align}
The structure of Eq. \eqref{43} shows explicitly that the Bumblebee fluctuations probe the conserved planar fermionic current. The temporal component $\delta B_0$ couples to the quasiparticle density, whereas $\delta B_1$ and $\delta B_2$ couple to the in-plane currents. The factors of $v_F$ accompanying the spatial vertices are required by the same anisotropic Dirac structure that appears in the kinetic term. Thus, these factors must be retained consistently in the polarization tensor.

According to the usual momentum-space Feynman rules, each interaction term $-g\bar{\psi}\mathcal{O}\psi\,\delta B$ contributes a vertex $-ig\mathcal{O}$. Hence, the three nonvanishing fermion-Bumblebee vertices are
\begin{align}
\delta B_0:-ig\gamma^0, \qquad \delta B_1:-igv_F\gamma^1 \qquad \mathrm{and} \qquad \delta B_2:-igv_F\gamma^2
\label{44}
\end{align}
The three vertex factors %share the same diagrammatic structure and 
differ only in the component of the external Bumblebee fluctuation. Introducing $\widetilde{\gamma}^{\,0}=\gamma^0$, $\widetilde{\gamma}^{\,1}=v_F\gamma^1$ and $\widetilde{\gamma}^{\,2}=v_F\gamma^2$, the corresponding momentum-space vertex can be written compactly as $\delta B_a:-ig\,\widetilde{\gamma}^{\,m}$ with $m=0,1,2$. Therefore, the diagram represents the elementary interaction between a planar Dirac quasiparticle and a Bumblebee fluctuation tangent to the sheet. The absence of an analogous vertex for $\delta B_3$ reflects the confinement of the fermionic current to the three-dimensional plane\footnote{In other words, there is no vertex proportional to $\delta B_3$. This result follows directly from the dimensional confinement of the quasiparticles and should not be interpreted as a gauge choice or as the neglect of a possible interaction. The fermionic field has no dynamical current in the normal direction; equivalently, the planar theory contains no derivative $\partial_3$ and therefore no current component capable of coupling minimally to $B_3$.}.

Consequently, this property has two immediate consequences. First, the one-loop fermionic effective action generated by the sheet is a functional only of $\delta B_0$, $\delta B_1$, and $\delta B_2$. Therefore, the corresponding polarization tensor carries exclusively planar indices, i.e., $\Pi^{ab}(p)$ with $a,b=0,1,2$. Second, the massive fluctuation associated with the curvature of the Bumblebee potential along the vacuum direction is not directly excited by a fermionic current localized on the sheet.% Its relevance must instead be determined from the bulk equations of motion, as will be done.

For the chosen orientation of the Lorentz-violating backgrounds, the fermionic loop retains the same structural form as the standard polarization loop of a massive planar Dirac material, with the electromagnetic coupling replaced by the Bumblebee coupling $g$. This equivalence concerns only the surface response. The complete scattering problem is not equivalent to the Maxwell case because the external field lines are Bumblebee-CFJ excitations whose bulk dispersion relations and polarization eigenvectors are modified by Lorentz violation.

%\section{The one-loop polarization tensor}
\section{Renormalized polarization and optical response}
\label{sectionIV}

In this section,  we do not introduce a new polarization-tensor calculation. We restate the standard one-loop vacuum-polarization tensor of massive Dirac fermions in $(2+1)$ dimensions, following well-known results such as those of Appelquist et al. \cite{appelquist1986spontaneous}, in order to fix the conventions used throughout the scattering problem. In the Bumblebee-CFJ setup, the parameters $b$ and $\kappa$ do not enter either the planar fermion propagator or the fermion-Bumblebee vertices. Therefore, the one-loop surface response has the same structure as that of a massive planar Dirac system, with the electromagnetic coupling replaced by the effective coupling $g$  \cite{Fialkovsky2011}. The only model-dependent ingredients that must be kept explicit at this stage are the coupling $g$ and the Fermi velocity $v_F$, since they enter the surface kernel used later in the Casimir interaction.

Within this framework, the one-loop polarization tensor is
\begin{align}
\Pi^{ab}(p_0,p_1,p_2)=ig^2N\,\int\,\frac{d^3q}{(2\pi)^3}\,\mathrm{tr}\left[S(q_0,q_1,q_2)\tilde{\gamma}^{\,a}S(q_0-p_0,q_1-p_1,q_2-p_2)\tilde{\gamma}^{\,b}\right],
\label{eq:polarization-general}
\end{align}
Thus, upon substituting Eq. \eqref{42} into Eq. \eqref{eq:polarization-general}, one has
\small{\begin{align}\nonumber
&\Pi^{ab}(p)=i g^{2}N \,\int\frac{d^3q}{(2\pi)^{3}}\times\\
&\mathrm{tr}\Big[\frac{\big(q_0\gamma^{0}+v_{F}q_{1}\gamma^{1}+v_{F}q_{2}\gamma^{2}+m\big)\tilde{\gamma}^{m}\big((q_{0}-p_{0})\gamma^{0}+v_{F}(q_{1}-p_{1})\gamma^{1}+v_{F}(q_{2}-p_{2})\gamma^{2}+m\big)\tilde{\gamma}^{n}}{[q_0^{2}-v_{F}^{2}(q_{1}^{2}+q_{2}^{2})-m^{2}][(q_{0}-p_{0})^{2}-v_{F}^{2}(q_{1}-p_{1})^{2}-v_{F}^{2}(q_{2}-p_{2})^{2}-m^{2}]}
\Big].
\label{eq:Pi-substitution}
\end{align}}

For graphene, the fermionic degeneracy factor is $N=4$, accounting for the two spin projections and the two inequivalent valleys $K$ and $K'$ \cite{Fialkovsky2011}. In the following, we consider ordinary graphene in the absence of a Haldane-like parity-breaking mass. The two valleys are described by the two inequivalent irreducible representations of the $(2+1)$-dimensional Clifford algebra, for which the trace of three Dirac matrices has opposite signs, $\operatorname{tr}_{K}\left(\gamma^{a}\gamma^{b}\gamma^{c}\right)=+2i\epsilon^{abc}$ and $\operatorname{tr}_{K'}\left(\gamma^{a}\gamma^{b}\gamma^{c}\right)=-2i\epsilon^{abc}$. Consequently, the parity-odd contributions proportional to
$im\epsilon^{abc}p_{c}$ cancel exactly after summing over the two valleys,
whereas the parity-even contributions add and acquire the full degeneracy
factor $N=4$ \cite{Fialkovsky2011}. Therefore, the polarization tensor of graphene contains only the parity-even contribution. Taking into account $\operatorname{tr}\left(\gamma^{a}\gamma^{b}\right)=2g^{ab}$ and $\operatorname{tr}\left(\gamma^{a}\gamma^{b}\gamma^{c}\gamma^{d}\right)=2\left(g^{ab}g^{cd}-g^{ac}g^{bd}+g^{ad}g^{bc}\right)$ with $g^{ab}=\operatorname{diag}(1,-1,-1)$, one obtains, for $a=b=0$,
\small{\begin{align}
\Pi^{00}(p)=&8i g^{2}\int\frac{d^3q}{(2\pi)^{3}}\Big[\frac{q_0(q_{0}-p_{0})+v_{F}^{2}q_{1}(q_{1}-p_{1})+v_{F}^{2}q_{2}(q_{2}-p_{2})+m^{2}}{[q_0^{2}-v_{F}^{2}(q_{1}^{2}+q_{2}^{2})-m^{2}][(q_{0}-p_{0})^{2}-v_{F}^{2}(q_{1}-p_{1})^{2}-v_{F}^{2}(q_{2}-p_{2})^{2}-m^{2}]}\Big].
\label{eq:Pi00-general}
\end{align}}

Meanwhile, let us consider $a=0$ and $b=i$ with $i=1,2$. In this case, by using $\tilde{\gamma}^{i}=v_{F}\gamma^{i}$, the parity-even Dirac trace gives
\begin{align}\nonumber
\Pi^{0i}(p)=&-8i g^{2} v_{F}^{2}\int\frac{d^3q}{(2\pi)^{3}}\Big[\frac{q_0(q_{i}-p_{i})+q_{i}(q_{0}-p_{0})}{[q_0^{2}-v_{F}^{2}(q_{1}^{2}+q_{2}^{2})-m^{2}][(q_{0}-p_{0})^{2}-v_{F}^{2}(q_{1}-p_{1})^{2}-v_{F}^{2}(q_{2}-p_{2})^{2}-m^{2}]}\Big].
%\label{eq:Pi0i-general}
\end{align}
Here, one highlights that $\Pi^{i0}(p)=\Pi^{0i}(p)$.

Finally, for $a=i$ and $b=j$, with $i,j=1,2$, one finds
\small{\begin{align}\nonumber
\Pi^{ij}(p)&=8i g^{2} v_{F}^{2}\int\frac{d^3q}{(2\pi)^{3}}\\
&\Bigg\{\frac{v_{F}^{2}\Big[q_{i}(q_{j}-p_{j})+q_{j}(q_{i}-p_{i})\Big]+\delta_{ij}\Big[q_0(q_{0}-p_{0})-v_{F}^{2}q_{1}(q_{1}-p_{1})-v_{F}^{2}q_{2}(q_{2}-p_{2})-m^{2}\Big]}{[q_0^{2}-v_{F}^{2}(q_{1}^{2}+q_{2}^{2})-m^{2}][(q_{0}-p_{0})^{2}-v_{F}^{2}(q_{1}-p_{1})^{2}-v_{F}^{2}(q_{2}-p_{2})^{2}-m^{2}]}\Bigg\}.
\label{eq:Pij-general}
\end{align}}

Applying the Feynman parametrization \cite{Feynman1949}, followed by the Pauli–Villars subtraction \cite{Pauli1949} and the momentum shift $q_i \to q_i-xp_i$, one obtains the well-known components of the renormalized polarization tensor \cite{appelquist1986spontaneous}, viz.,
\begin{align}\nonumber
\Pi^{00}_{R}(p)=&\frac{g^{2}\left(p_{1}^{2}+p_{2}^{2}\right)}{2\pi\left[p_{0}^{2}-v_{F}^{2}\left(p_{1}^{2}+p_{2}^{2}\right)\right]^{3/2}}\Bigg\{-2|m|\sqrt{p_{0}^{2}-v_{F}^{2}\left(p_{1}^{2}+p_{2}^{2}\right)}+\left[4m^{2}+p_{0}^{2}-v_{F}^{2}\left(p_{1}^{2}+p_{2}^{2}\right)\right]\times\\
&\operatorname{arctanh}\left[\frac{\sqrt{p_{0}^{2}-v_{F}^{2}\left(p_{1}^{2}+p_{2}^{2}\right)}}{2|m|}\right]\Bigg\},
\label{eq:Pi00-ren-final}
\end{align}
\begin{align}\nonumber
    \Pi_{R}^{0i}(p)=&-\frac{g^2 p_0p_i}{2\pi\left[p_0^2-v_F^2(p_1^2+p_2^2)\right]^{3/2}}\Bigg\{-2\vert m\vert\sqrt{p_0^2-v_F^2(p_1^2+p_2^2)}+[4m^2+p_0^2-v_F^2(p_1^2+p_2^2)]\times\\
    &\mathrm{arctanh}\,\left[\frac{\sqrt{p_0^2-v_F^2(p_1^2+p_2^2)}}{2\vert m \vert}\right]\Bigg\},
    \label{eq:PV-Pi0i-ren2}
\end{align}
and
\begin{align}\nonumber
    \Pi_{R}^{ij}(p)=&\frac{g^2}{2\pi\left[p_0^2-v_F^2(p_1^2+p_2^2)\right]^{3/2}}\Bigg\{-2\vert m \vert\sqrt{p_0^2-v_F^2(p_1^2+p_2^2)}+[4m^2+p_0^2-v_F^2(p_1^2+p_2^2)]\times\\
    &\mathrm{arctanh}\left[\frac{\sqrt{p_0^2-v_F^2(p_1^2+p_2^2)}}{2\vert m \vert}\right]\Bigg\}[\delta_{ij}[p_0^2-v_F^2(p_1^2+p_2^2)]+v_F^2p_ip_j]
    \label{eq:PV-Piij-ren2}
\end{align}
with $i,j=1,2$. For more details on the corresponding massive $(2+1)$-dimensional
vacuum polarization, see the Refs. \cite{gusynin19962H,appelquist1986spontaneous}\footnote{In Ref. \cite{gusynin19962H,appelquist1986spontaneous}, the authors adopt $v_F=1$.}. Indeed, note that the renormalized polarization tensor in Eqs. \eqref{eq:Pi00-ren-final}-\eqref{eq:PV-Piij-ren2} corresponds to the standard one-loop response of massive Dirac fermions in $(2+1)$ dimensions. For the Bumblebee-CFJ background configuration, neither $b$ nor $\kappa$ enters the fermionic propagator or the interaction vertices. Therefore, the surface response coincides with the well-known massive Dirac result, with the electromagnetic coupling replaced by the effective Bumblebee coupling $g$ \cite{gusynin19962H}. Naturally, these
expressions provide the finite-momentum response required in the next section for the construction of the surface kernel and the Casimir interaction.% No new property of the planar polarization tensor is claimed here. These expressions are included only to fix our conventions and to provide the nonlocal surface response required for the Casimir problem.

Using the polarization tensor announced in Eqs. \eqref{eq:Pi00-ren-final}-\eqref{eq:PV-Piij-ren2}, one can proceed to deriving the optical conductivity. In this case, upon using the retarded frequency $z=\omega+i0^+$ and the gauge-invariant Kubo relation \cite{sharapov2005effects} for the conductivity, i.e., 
\begin{align}
    \sigma^{ij}(z)=-\frac{i}{z}\left[\Pi^{ij}_{R}(z,\mathbf{0})-\Pi^{ij}_{R}(0,\mathbf{0})\right],
\end{align}
with $z=\omega+i0^{+}$. Within the Pauli-Villars prescription adopted, $\Pi^{ij}_{R}(0,\mathbf{0})=0$, and the diagonal components reduce to the usual optical response of a gapped Dirac system
\begin{align}
    \sigma^{11}(z)=\sigma^{22}(z)=\frac{i g^{2}}{4\pi}\left[\frac{4|m|}{z}-\left(1+\frac{4m^{2}}{z^{2}}\right)\ln\left(\frac{2|m|+z}{2|m|-z}\right)\right].
    \label{cond111}
\end{align}
The absorptive response starts at the interband threshold $\omega=2|m|$, while the gapless limit approaches $g^{2}/4$, in agreement with the standard Dirac conductivity \cite{klimchitskaya2017conductivity1}. These established results will not be discussed further. For more details on conductivity, see Ref. \cite{klimchitskaya2017conductivity1}.

%%%%%%%%%%%%%%%%%%%%%%%%%%%%%%%%%%%%%%%%%%%

\section{The Casimir interaction}
\label{sectionVI}

Now, let us apply the renormalized polarization tensor to the Casimir interaction between two identical, parallel Dirac sheets at $x^3=0$ and $x^3=a$, with $a>0$. The Bumblebee-CFJ background is homogeneous in the intervening region and on both sides of the sheets. The material input is supplied by Eqs. \eqref{eq:Pi00-ren-final}, \eqref{eq:PV-Pi0i-ren2}, and \eqref{eq:PV-Piij-ren2}, while the interaction is obtained from the scattering formulation of the Lifshitz energy \cite{Fialkovsky2018,casrahi,cas-fukushima}. Throughout the calculation, the full dependence on the planar momentum is retained.% The interpretation of the resulting force as an electromagnetic contribution follows the identification of the transverse Bumblebee excitation with the photon discussed in Sec. \ref{sectionV}.

\begin{figure}[!ht]
\centering
\includegraphics[width=0.45\textwidth]{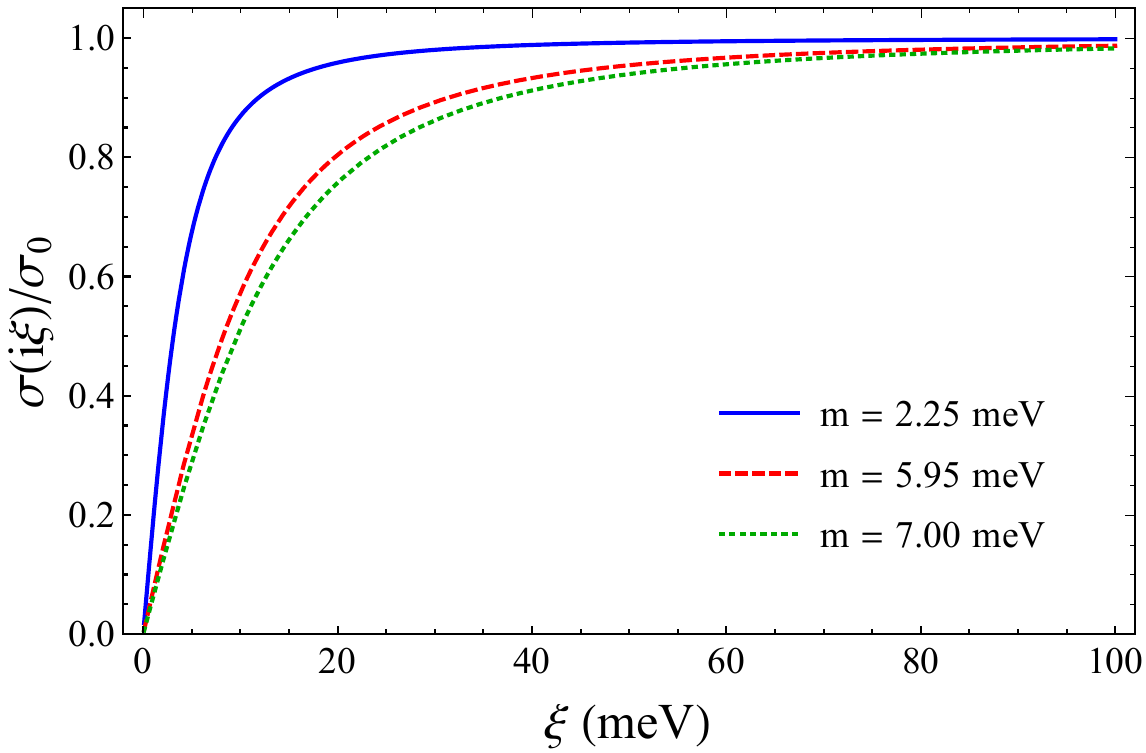}
\caption{Conductivity on the imaginary-frequency axis, normalized by $\sigma_0=g^2/4$, for $m=2.25$, $5.95$, and $7.00\,\mathrm{meV}$. The gap suppresses the low-frequency response, and all curves approach the gapless value at large $\xi$.}
\label{fig:imag-conductivity}
\end{figure}

On the imaginary-frequency axis, we write $p_0=i\xi$, with $\xi>0$, and $\mathbf{k}=(p_1,p_2)$, with $k^2=p_1^2+p_2^2$. Analytic continuation of Eq. \eqref{cond111} gives, for $m\ne0$,
\begin{align}
\sigma(i\xi)=\frac{g^2}{2\pi}
\left[\frac{2|m|}{\xi}
+\left(1-\frac{4m^2}{\xi^2}\right)
\arctan\left(\frac{\xi}{2|m|}\right)\right].
\label{eq:imag-conductivity}
\end{align}
This response is real and nonnegative. Its low-frequency expansion is
\begin{align}
\sigma(i\xi)=\frac{g^2\xi}{3\pi|m|}-\frac{g^2\xi^3}{30\pi|m|^3}+\mathcal{O}\left(\frac{g^2\xi^5}{|m|^5}\right) \qquad \mathrm{with} \qquad \xi\ll|m|,
\label{eq:imag-expansion}
\end{align}
whereas the gapless value is $\sigma_0=g^2/4$. Fig. \ref{fig:imag-conductivity} shows the suppression of the response at small $\xi$ and its approach to $\sigma_0$ when the frequency exceeds the mass scale. This optical result describes the $k=0$ limit.

Meanwhile, the bulk modes that couple to this response follow from the quadratic expansion of the potential announced in Eq. \eqref{eq:SB}, viz., $X=2b_\mu\delta B^\mu+\delta B_\mu\delta B^\mu$, which leads us to
\begin{align}
V(X)&=V(0)+2V''(0)(b_\mu\delta B^\mu)^2+\mathcal{O}((\delta B)^3).
\label{eq:potential-expansion}
\end{align}
Here $\delta B_{\mu\nu}=\partial_\mu\delta B_\nu-\partial_\nu\delta B_\mu$ denotes the field-strength tensor of the Bumblebee fluctuation. The linearized equation of motion for $\delta B_\mu$, sourced by the induced sheet currents evaluated to first order in the fluctuation, reads
\begin{align}
J_B^\nu=\partial_\mu\delta B^{\mu\nu}+(\kappa_{\mathrm{AF}})_\mu\epsilon^{\mu\nu\rho\sigma}\delta B_{\rho\sigma}-4V''(0)b^\nu(b_\rho\delta B^\rho).
\label{eq:bulk-field}
\end{align}
The current is conserved as a distribution and has no normal component, $J_B^3=0$. Therefore, taking the divergence of Eq. \eqref{eq:bulk-field}, and using the antisymmetry of the field strength and the Bianchi identity, yields $(b^\nu\partial_\nu)(b_\rho\delta B^\rho)=0$. For the nonzero normal background in Eq. \eqref{eq:backgrounds}, this condition reduces to $\partial_3\delta B^3=0$\footnote{Since $\delta B^3$ is continuous across the sheets and no localized normal current is present, the same constant value is fixed in the region between the sheets. The exterior decay condition then sets this constant to zero everywhere.}. Exterior decay, together with the absence of an independent source for the normal fluctuation, then requires $\delta B^3=0$. Thus, the $\nu=3$ component of Eq. \eqref{eq:bulk-field} further gives $\partial_3(\partial_a\delta B^a)=0$, so the same exterior condition fixes $\partial_a\delta B^a=0$. Thereby, the normal fluctuation is excluded by the field equations and boundary conditions, and the quadratic potential does not contribute to the propagating sector. Consequently, the parameter $b$ does not contribute to this interaction, while the CFJ term remains in the tangential equations.

For the Euclidean momentum $(\xi,k_1,k_2)$, an orthonormal basis of the transverse subspace is
\begin{align}
q=\sqrt{\xi^2+k^2},\qquad e_{\mathrm{TM}}^a=\left(\frac{k}{q},-\frac{\xi k_1}{qk},-\frac{\xi k_2}{qk}\right) \qquad \mathrm{and} \qquad e_{\mathrm{TE}}^a=\left(0,-\frac{k_2}{k},\frac{k_1}{k}\right).
\label{eq:transverse-basis}
\end{align}
The case $k=0$ is defined by continuity. Therefore, writing $\delta B_E^a=e_{\mathrm{TM}}^a\beta_{\mathrm{TM}}+e_{\mathrm{TE}}^a\beta_{\mathrm{TE}}$ and $\boldsymbol{\beta}=(\beta_{\mathrm{TM}},\beta_{\mathrm{TE}})^t$, projection of the bulk equation gives
\begin{align}
\left[-\partial_3^2\mathbb{I}
+\begin{pmatrix}q^2&2\kappa q\\-2\kappa q&q^2\end{pmatrix}\right]
\boldsymbol{\beta}=0.
\label{eq:projected-bulk}
\end{align}
The factor $2\kappa$ is fixed by the CFJ normalization in Eq. \eqref{eq:SB}. Reversing the TE basis vector changes the signs of both off-diagonal entries without affecting the interaction. For a solution proportional to $e^{-\rho x^3}$, the characteristic equation and the decaying branches are
\begin{align}
(q^2-\rho^2)^2+4\kappa^2q^2=0,\qquad \mathrm{and} \qquad
\rho_\pm=\sqrt{q^2\pm2i\kappa q},
\label{eq:euclidean-roots}
\end{align}
with $\operatorname{Re}\rho_\pm>0$.

Although these attenuation constants are complex conjugates, the corresponding real-frequency dispersion satisfies
\begin{align}
(\omega^2-k^2-k_3^2)^2-4\kappa^2(\omega^2-k^2)&=0,
\end{align}
i.e., 
\begin{align}
\omega_\pm^2&=k^2+\left(\sqrt{k_3^2+\kappa^2}\pm|\kappa|\right)^2.
\label{eq:real-dispersion}
\end{align}
Hence, the complex Euclidean roots do not signal a real-frequency instability for this space-like background. The same conjugate structure occurs for a normal axion gradient \cite{cas-fukushima}, whose magnitude corresponds to $2|\kappa|$ in the present normalization.

A real matrix representation is obtained by defining
\begin{align}
u=\sqrt{\frac{q\sqrt{q^2+4\kappa^2}+q^2}{2}},\qquad v=\frac{\kappa q}{u}, \qquad \mathbb{Q}=\begin{pmatrix}u&v\\-v&u\end{pmatrix}, \qquad \mathrm{and} \qquad
\mathbb{Q}^2=\begin{pmatrix}q^2&2\kappa q\\-2\kappa q&q^2\end{pmatrix}.
\label{eq:Q-definition}
\end{align}
Here, $u=\operatorname{Re}\rho_+$ and $v=\operatorname{Im}\rho_+$, with $u\ge q$, $u^2-v^2=q^2$, and $uv=\kappa q$. Thereby, The translation across the separation is %therefore described by
\begin{align}
\mathbb{U}(a)%=e^{-a\mathbb{Q}}
=e^{-au}\begin{pmatrix}\cos(av)&-\sin(av)\\\sin(av)&\cos(av)\end{pmatrix}.
\label{eq:translation}
\end{align}
This expression incorporates both attenuation and mixing of the TE and TM amplitudes.% It reduces to $e^{-aq}\mathbb{I}$ when $\kappa=0$.

The surface-response kernel is obtained by projecting the Euclidean continuation of Eqs. \eqref{eq:Pi00-ren-final}, \eqref{eq:PV-Pi0i-ren2}, and \eqref{eq:PV-Piij-ren2} onto the same TM/TE basis. In this basis, the response of the Dirac sheet is diagonal and can be represented by the surface kernel $\Sigma$, whose eigenvalues are determined by the longitudinal and transverse conductivities at finite momentum, $\sigma_L(i\xi,k)$ and $\sigma_T(i\xi,k)$, respectively. Thus,
%Naturally, the surface kernel is obtained by projecting the Euclidean continuation of Eqs.\eqref{eq:Pi00-ren-final}, \eqref{eq:PV-Pi0i-ren2}, and \eqref{eq:PV-Piij-ren2} onto the same basis. Thus, one obtains in terms of the longitudinal and transverse conductivities at finite momentum, $\sigma_L(i\xi,k)$ and $\sigma_T(i\xi,k)$, i.e., 
\begin{align}
\boldsymbol{\Sigma}=\begin{pmatrix}\lambda_{\mathrm{TM}}&0\\0&\lambda_{\mathrm{TE}}\end{pmatrix},
\label{eq:surface-kernel}
\end{align}
with $\lambda_{\mathrm{TM}}=\frac{q^2}{\xi}\sigma_L(i\xi,k)$ and $\lambda_{\mathrm{TE}}=\xi\sigma_T(i\xi,k)$, where the two eigenvalues are
\begin{align}
\begin{pmatrix}\lambda_{\mathrm{TM}}\\\lambda_{\mathrm{TE}}\end{pmatrix}
=\frac{2g^2}{\pi}
\begin{pmatrix}\xi^2+k^2\\\xi^2+v_F^2k^2\end{pmatrix}
\int_0^1 dx\,
\frac{x(1-x)}{\sqrt{m^2+x(1-x)(\xi^2+v_F^2k^2)}}.
\label{eq:surface-eigenvalues}
\end{align}
In the TM contraction $e_{\mathrm{TM}}^a\Pi_E^{ab}e_{\mathrm{TM}}^b$, the temporal, mixed, and spatial terms contribute the factors $k^4/q^2$, $2\xi^2k^2/q^2$, and $\xi^4/q^2$, respectively. Their sum is $q^2$, which accounts for the TM prefactor in Eq.~\eqref{eq:surface-eigenvalues}. Orthogonality to $\mathbf{k}$ eliminates the mixed surface eigenvalue. At $k=0$, both conductivities reduce to Eq. \eqref{eq:imag-conductivity}; at finite $k$, Eq.~\eqref{eq:surface-eigenvalues} supplies the spatial dispersion required in the interaction energy.

For a sheet at $x^3=0$, the bulk equation including its induced current takes the form
\begin{align}
\left[-\partial_3^2\mathbb{I}+\mathbb{Q}^2
+\delta(x^3)\boldsymbol{\Sigma}\right]\boldsymbol{\beta}(x^3)=0.
\label{eq:sheet-wave}
\end{align}
Continuity of the field and integration across the sheet give $\boldsymbol{\beta}(0^+)=\boldsymbol{\beta}(0^-)$ and $\partial_3\boldsymbol{\beta}(0^+)-\partial_3\boldsymbol{\beta}(0^-)
=\boldsymbol{\Sigma}\boldsymbol{\beta}(0)$. The uniform, normally oriented CFJ background adds no surface delta function, since its contribution in Eq. \eqref{eq:projected-bulk} contains only tangential derivatives. To solve the matching conditions, write the analytically continued scattering field as
\begin{align}
\boldsymbol{\beta}(x^3<0)=e^{-x^3\mathbb{Q}}\boldsymbol{\beta}_{\mathrm{in}}
+e^{x^3\mathbb{Q}}\boldsymbol{\beta}_{\mathrm{r}} \qquad \mathrm{and} \qquad \boldsymbol{\beta}(x^3>0)=e^{-x^3\mathbb{Q}}\boldsymbol{\beta}_{\mathrm{t}}.
\label{eq:scattering-fields}
\end{align}
The growing incident factor is the imaginary-frequency continuation of an incoming wave. Therefore, applying the matching conditions derived from Eq. \eqref{eq:sheet-wave} to the scattering ansatz \eqref{eq:scattering-fields} gives
%The growing incident factor is the imaginary-frequency continuation of an incoming wave. Substitution into Eq. \eqref{eq:matching} yields 
$\boldsymbol{\beta}_{\mathrm{t}}=\boldsymbol{\beta}_{\mathrm{in}}+\boldsymbol{\beta}_{\mathrm{r}}$ and $-2\mathbb{Q}\boldsymbol{\beta}_{\mathrm{r}}=\boldsymbol{\Sigma}(\boldsymbol{\beta}_{\mathrm{in}}+\boldsymbol{\beta}_{\mathrm{r}})$, with $\boldsymbol{\beta}_{\mathrm{r}}=\mathbb{R}\boldsymbol{\beta}_{\mathrm{in}}$ and $\mathbb{R}=-(2\mathbb{Q}+\boldsymbol{\Sigma})^{-1}\boldsymbol{\Sigma}$.

Meanwhile, the matrix order follows directly from the matching conditions, since $\mathbb{Q}$ and $\boldsymbol{\Sigma}$ do not commute. Explicitly, one has
\begin{align}
\mathbb{R}=\frac{1}{(2u+\lambda_{\mathrm{TM}})(2u+\lambda_{\mathrm{TE}})+4v^2}\begin{pmatrix}
-\lambda_{\mathrm{TM}}(2u+\lambda_{\mathrm{TE}})&2v\lambda_{\mathrm{TE}}\\ -2v\lambda_{\mathrm{TM}}&-\lambda_{\mathrm{TE}}(2u+\lambda_{\mathrm{TM}})
\end{pmatrix}.
\label{eq:reflection-explicit}
\end{align}
Therefore, the CFJ contribution to the bulk propagation operator couples the TE and TM amplitudes. This coupling appears in both the translation and reflection matrices.

Using the fixed tangential basis of Eq. \eqref{eq:transverse-basis} on both sheets, reflection from either side is represented by the same $\mathbb{R}$. In the limit $\kappa\to 0$, one has $\left.\mathbb{R}\right|_{\kappa=0}=\operatorname{diag}(-r_{\mathrm{TM}},r_{\mathrm{TE}})$, with
\begin{align}
r_{\mathrm{TM}}=\frac{q\sigma_L}{2\xi+q\sigma_L}%=\frac{q\Pi_E^{00}}{2k^2+q\Pi_E^{00}} 
\qquad \mathrm{and} \qquad
r_{\mathrm{TE}}=-\frac{\xi\sigma_T}{2q+\xi\sigma_T}.
\label{eq:maxwell-reflections}
\end{align}
The TM sign reflects the convention for the projected potential amplitude and cancels in a round trip. The normalization is fixed by the bulk kinetic term $-B_{\mu\nu}B^{\mu\nu}/4$, and the degeneracy $N=4$ is already included in the surface response.

After reflection at both sheets and propagation across the gap in each direction, the fluctuation acquires the round-trip matrix $\mathbb{M}(a)=\mathbb{R}\mathbb{U}(a)\mathbb{R}\mathbb{U}(a)$. Subtracting the homogeneous bulk vacuum and the self-energies of the isolated sheets, the Gaussian determinant gives the interaction energy per unit area, i.e., 
\begin{align}
\mathcal{E}(a)=\frac{1}{4\pi^2}\int_0^\infty\,d\xi\int_0^\infty k\,dk\,\ln\det\left[\mathbb{I}-\mathbb{R}e^{-a\mathbb{Q}}\mathbb{R}e^{-a\mathbb{Q}}\right].
\label{eq:energy}
\end{align}
%The Gaussian factor $1/2$ is compensated by combining positive and negative frequencies. All orders of multiple reflection are retained, as is evident from
%\begin{align}
%\ln\det(\mathbb{I}-\mathbb{M})
%=-\sum_{n=1}^{\infty}\frac{1}{n}\operatorname{tr}(\mathbb{M}^n),
%\label{eq:multiple-reflections}
%\end{align}
%where $n$ counts complete round trips. The material response remains at quadratic, one-loop order.
For numerical purposes, the identity $\det(\mathbb{I}-\mathbb{A}^2)=1-[\operatorname{tr}\mathbb{A}]^2+2\det\mathbb{A}+(\det\mathbb{A})^2$, applied to $\mathbb{A}=\mathbb{R}\mathbb{U}$, gives the real expression
\begin{align}
\begin{aligned}
\det[\mathbb{I}-\mathbb{R}\mathbb{U}\mathbb{R}\mathbb{U}]=&1-e^{-2au}\Big\{
\big[(R_{11}+R_{22})\cos(av)+(R_{12}-R_{21})\sin(av)\big]^2
-2\det\mathbb{R}\Big\}+e^{-4au}(\det\mathbb{R})^2,
\end{aligned}
\label{eq:real-determinant}
\end{align}
with
\begin{align}
\det\mathbb{R}=\frac{\lambda_{\mathrm{TM}}\lambda_{\mathrm{TE}}}
{(2u+\lambda_{\mathrm{TM}})(2u+\lambda_{\mathrm{TE}})+4v^2}.
\label{eq:reflection-determinant}
\end{align}
The logarithm is chosen continuously from $a\to\infty$, where the interaction determinant approaches unity. The pressure follows by differentiating at fixed dimensional parameters,
\begin{align}
\mathcal{P}(a)=-\left.\frac{\partial\mathcal{E}}{\partial a}\right|_{m,\kappa,g,v_F},
\label{eq:pressure-definition}
\end{align}
i.e., 
%Since $\mathbb{R}$ and $\mathbb{Q}$ are independent of $a$ at fixed $\xi$ and $k$, while $\partial_a\mathbb{U}=-\mathbb{Q}\mathbb{U}$, this derivative can be evaluated directly as
\begin{align}
\mathcal{P}(a)=-\frac{1}{4\pi^2}\int_0^\infty d\xi\int_0^\infty k\,dk\,
\operatorname{tr}\!\left\{(\mathbb{I}-\mathbb{M})^{-1}
\left[\mathbb{R}\mathbb{Q}\mathbb{U}\mathbb{R}\mathbb{U}
+\mathbb{R}\mathbb{U}\mathbb{R}\mathbb{Q}\mathbb{U}\right]\right\}.
\label{eq:pressure-integral}
\end{align}
Negative pressure denotes attraction, and positive pressure denotes repulsion.

Several limits clarify the physical content and normalization of these expressions. When $\kappa=0$, the two polarizations decouple and Eq.~\eqref{eq:energy} reduces to
\begin{align}
\left.\mathcal{E}(a)\right|_{\kappa=0}
=\frac{1}{4\pi^2}\int_0^\infty d\xi\int_0^\infty k\,dk\,
\left[\ln(1-r_{\mathrm{TM}}^2e^{-2aq})
+\ln(1-r_{\mathrm{TE}}^2e^{-2aq})\right].
\label{eq:maxwell-energy}
\end{align}

The nonnegative surface eigenvalues imply $0\le r_{\mathrm{TM}}<1$ and $-1<r_{\mathrm{TE}}\le0$, so
\begin{align}
\left.\mathcal{P}(a)\right|_{\kappa=0}
=-\frac{1}{2\pi^2}\int_0^\infty d\xi\int_0^\infty k\,dk\,q
\sum_{\alpha=\mathrm{TM},\mathrm{TE}}
\frac{r_\alpha^2e^{-2aq}}{1-r_\alpha^2e^{-2aq}}<0
\label{eq:maxwell-pressure}
\end{align}
for a nonzero sheet response.

For perfectly reflecting sheets, $\lambda_{\mathrm{TM}},\lambda_{\mathrm{TE}}\to\infty$ and $\mathbb{R}\to-\mathbb{I}$. The determinant becomes
$\det(\mathbb{I}-\mathbb{U}^2)
%&=1-2e^{-2au}\cos(2av)+e^{-4au}\\
=(1-e^{-2a\rho_+})(1-e^{-2a\rho_-})$. Setting also $\kappa=0$ and using $\xi=q\cos\theta$, $k=q\sin\theta$, with $0\le\theta\le\pi/2$, yields
\begin{align}
\begin{aligned}
\mathcal{E}_{\mathrm{ideal}}(a)=-\frac{\pi^2}{720a^3} \qquad \mathrm{and} \qquad \mathcal{P}_{\mathrm{ideal}}(a)=-\frac{\pi^2}{240a^4}.
\end{aligned}
\label{eq:perfect-maxwell}
\end{align}
This recovers the standard normalization and the two-polarization multiplicity.

For gapless sheets, the Feynman-parameter integral in Eq.~\eqref{eq:surface-eigenvalues} simplifies to
\begin{align}
\int_0^1 dx\,\frac{x(1-x)}{\sqrt{x(1-x)(\xi^2+v_F^2k^2)}}
=\frac{\pi}{8\sqrt{\xi^2+v_F^2k^2}}.
\label{eq:gapless-integral}
\end{align}
Thus, one obtains%and hence
\begin{align}
\sigma_L(i\xi,k)=\frac{g^2\xi}{4\sqrt{\xi^2+v_F^2k^2}}\qquad
 \mathrm{and} \qquad \sigma_T(i\xi,k)=\frac{g^2\sqrt{\xi^2+v_F^2k^2}}{4\xi}.
\label{eq:gapless-conductivities}
\end{align}
At $\kappa=0$, introduce $t=\xi/q\in[0,1]$, so that $k=q\sqrt{1-t^2}$ and $d\xi\,k\,dk=q^2\,dq\,dt$. The reflection amplitudes depend only on $t$, i.e., 
\begin{align}
\begin{aligned}
r_{\mathrm{TM}}(t)=\frac{g^2}{8\sqrt{v_F^2+(1-v_F^2)t^2}+g^2} \qquad \mathrm{and} \qquad
r_{\mathrm{TE}}(t)=-\frac{g^2\sqrt{v_F^2+(1-v_F^2)t^2}}
{8+g^2\sqrt{v_F^2+(1-v_F^2)t^2}}.
\end{aligned}
\label{eq:gapless-reflections}
\end{align}
Expanding the logarithm and integrating over the radial variable gives
\begin{align}
\int_0^\infty q^2\,dq\,\ln(1-r^2e^{-2aq})
=-\frac{1}{4a^3}\sum_{n=1}^{\infty}\frac{r^{2n}}{n^4}
=-\frac{\operatorname{Li}_4(r^2)}{4a^3}.
\label{eq:radial-polylog}
\end{align}

Consequently, the energy and pressure reduce to\footnote{Here $\operatorname{Li}4(z)$ denotes the polylogarithm of order four, defined for $|z|\leq1$ by $\operatorname{Li}4(z)=\sum{n=1}^{\infty}z^n/n^4$. Its argument is $z=r\alpha^2(t)$, with $\alpha=\mathrm{TM},\mathrm{TE}$.}
\begin{align}
\left.\mathcal{E}(a)\right|_{m=\kappa=0}
=-\frac{1}{16\pi^2a^3}\int_0^1dt\,
\left[\operatorname{Li}_4(r_{\mathrm{TM}}^2(t))
+\operatorname{Li}_4(r_{\mathrm{TE}}^2(t))\right],
\end{align}
and
\begin{align}
\left.\mathcal{P}(a)\right|_{m=\kappa=0}&=\frac{3\mathcal{E}(a)}{a}.
\label{eq:gapless-energy}
\end{align}
The $a^{-3}$ energy scaling follows from the absence of a mass or CFJ length scale. Its coefficient nevertheless depends on $v_F$ through spatial dispersion, despite the cancellation of $v_F$ in the optical conductivity.

For gapped sheets at $\kappa=0$ and $|m|a\gg1$, the dominant momenta obey $\sqrt{\xi^2+v_F^2k^2}\ll|m|$ at fixed $g$ and $0<v_F\le1$. Expanding the surface response gives
\begin{align}
\frac{2g^2}{\pi}\int_0^1dx\,\frac{x(1-x)}{\sqrt{m^2+x(1-x)(\xi^2+v_F^2k^2)}}=\frac{g^2}{3\pi|m|}+\mathcal{O}\left(\frac{g^2(\xi^2+v_F^2k^2)}{|m|^3}\right),
\label{eq:gapped-expansion}
\end{align}
which implies
\begin{align}
r_{\mathrm{TM}}\simeq\frac{g^2q}{6\pi|m|},\qquad
r_{\mathrm{TE}}\simeq-\frac{g^2(\xi^2+v_F^2k^2)}{6\pi|m|q}.
\label{eq:gapped-reflections}
\end{align}
Naturally, both amplitudes become small at large separation. Keeping the leading term in the logarithms of Eq. \eqref{eq:maxwell-energy} yields
\begin{align}
\mathcal{E}(a)\simeq-\frac{g^4}{144\pi^4m^2}
\int_0^\infty q^4\,dq\,e^{-2aq}
\int_0^1dt\,\left[1+\big(v_F^2+(1-v_F^2)t^2\big)^2\right].
\label{eq:gapped-energy-integral}
\end{align}
Using the identities
\begin{align}
\int_0^\infty q^4e^{-2aq}\,dq=\frac{3}{4a^5} \qquad \mathrm{and} \qquad 
\int_0^1dt\,\left[1+\big(v_F^2+(1-v_F^2)t^2\big)^2\right]=\frac{18+4v_F^2+8v_F^4}{15},
\label{eq:gapped-integrals}
\end{align}
one obtains
\begin{align}
\left.\mathcal{E}(a)\right|_{\kappa=0}
\simeq-\frac{g^4(9+2v_F^2+4v_F^4)}{1440\pi^4m^2a^5},\qquad
\left.\mathcal{P}(a)\right|_{\kappa=0}
\simeq-\frac{g^4(9+2v_F^2+4v_F^4)}{288\pi^4m^2a^6}.
\label{eq:gapped-asymptote}
\end{align}
The suppression of low-frequency fluctuations by the gap thus changes the large-distance energy scaling from $a^{-3}$ to $a^{-5}$. These asymptotic powers apply to Maxwell propagation; a nonzero $\kappa$ introduces an additional propagation scale.

Furthermore, the dependence on the Lorentz-violating background is constrained by a further symmetry. Under $\kappa\to-\kappa$, $u$ remains unchanged and $v\to-v$. Both $\mathbb{R}$ and $\mathbb{U}$ consequently transform by conjugation with $\mathbb{S}=\operatorname{diag}(1,-1)$, leaving the interaction determinant invariant. Together with $\delta B^3=0$, this gives
\begin{align}
\mathcal{E}(a;\kappa)=\mathcal{E}(a;-\kappa) \qquad \mathrm{and} \qquad \frac{\partial\mathcal{E}}{\partial b}=0
\label{eq:background-dependence}
\end{align}
in the fluctuation sector. 

The CFJ contribution modifies both the attenuation of the bulk modes and the coupling between their polarizations. To determine its effect on the interaction, we numerically evaluate the energy integral in Eq.~\eqref{eq:energy} and the pressure integral in Eq. \eqref{eq:pressure-integral}\footnote{For the numerical integration, we introduce $s=qa$, with $q=\sqrt{\xi^2+k^2}$, and $t=\xi/q\in[0,1]$, such that $d\xi\,k\,dk=q^2\,dq\,dt$. The radial integration is truncated at $s_{\max}=40$, where the translation factor is exponentially suppressed. Convergence was verified by comparing Gauss-Legendre quadratures with 160 and 240 nodes per variable. The gapless and ideal-conductor limits provide independent checks, while near $q=0$ the Feynman-parameter representation of the surface response, or its low-momentum expansion, is used to avoid numerical cancellation.}, taking $g^2=4\pi/137$ and $v_F=1/300$. The material input to both calculations consists of the surface eigenvalues in Eq.~\eqref{eq:surface-eigenvalues}, which enter the reflection matrix through Eq.~\eqref{eq:reflection-explicit}. Bulk propagation is specified by Eqs.~\eqref{eq:Q-definition} and \eqref{eq:translation}. These expressions retain the full spatial dispersion of the sheet response and all orders of multiple reflection.

\begin{figure}[!ht]
    \centering
    \includegraphics[width=0.45\linewidth]{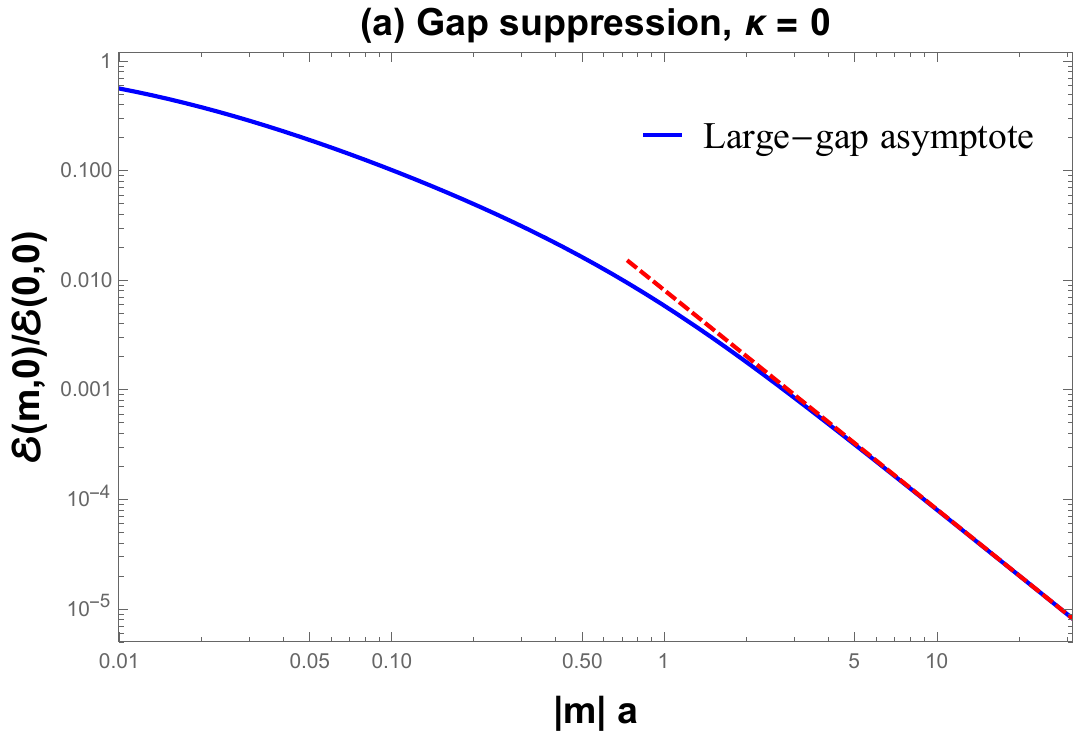}
    \includegraphics[width=0.45\linewidth]{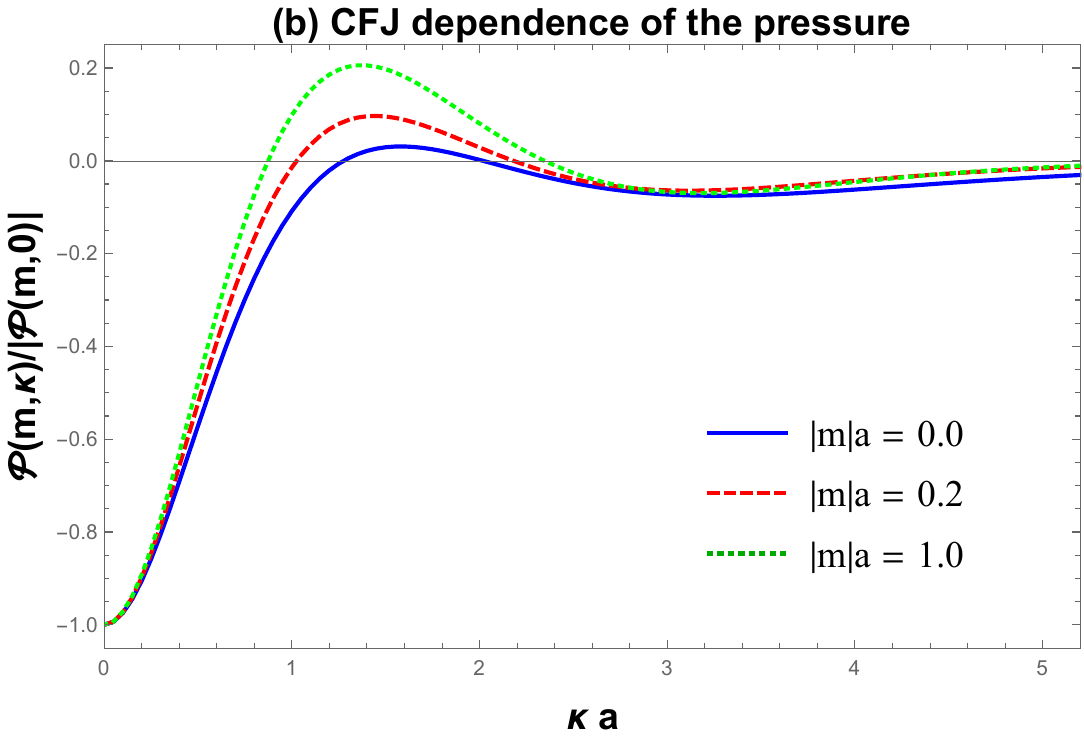}
    \caption{Casimir interaction between two identical sheets with $g^2=4\pi/137$ and $v_F=1/300$. (a) Suppression of the energy magnitude by the gap at $\kappa=0$; the dashed curve is the large-distance asymptote \eqref{eq:gapped-asymptote}. (b) Pressure vs. $\kappa a$ at the indicated values of $|m|a$.}
\label{fig:casimir}
\end{figure}

Fig. \ref{fig:casimir}(a) is constructed by evaluating the energy at $\kappa=0$, for which Eq. \eqref{eq:energy} reduces to Eq. \eqref{eq:maxwell-energy}. The resulting values are plotted as $\mathcal{E}(a;m,0)/\mathcal{E}(a;0,0)$ against $|m|a$, with the gapless reference energy calculated from Eq. \eqref{eq:gapless-energy}. The dashed curve represents the large-separation approximation in Eq. \eqref{eq:gapped-asymptote}, normalized by the same reference energy. Meanwhile, Fig. \ref{fig:casimir}(b) is obtained by numerically evaluating Eq. \eqref{eq:pressure-integral} as a function of $\kappa a$ for the indicated values of $|m|a$. The plotted quantity is $\mathcal{P}(a;m,\kappa)/|\mathcal{P}(a;m,0)|$, where the denominator is calculated from Eq.~\eqref{eq:maxwell-pressure} at the same mass and separation. The pressure expression follows from differentiation at fixed dimensional $m$, $\kappa$, $g$, and $v_F$, as prescribed in Eq.~\eqref{eq:pressure-definition}. Therefore, the positive values identify repulsive intervals. Table~\ref{tab:casimir} lists the corresponding dimensionless energy and pressure obtained from Eqs. \eqref{eq:energy} and \eqref{eq:pressure-integral}.

Furthermore, Fig. \ref{fig:casimir}(a) shows the progressive reduction of the interaction as the gap opens and the approach to the $m^{-2}a^{-5}$ asymptote. For the chosen parameters, the gapless Maxwell value is $a^3\mathcal{E}=-6.65231\times10^{-5}$. Fig. \ref{fig:casimir}(b) exhibits a different effect: increasing $|\kappa|a$ initially weakens the attraction, produces a repulsive interval, and restores attraction within the displayed range. The repulsive intervals found in this scan are
\begin{align}
\begin{aligned}
1.276&\lesssim\kappa a\lesssim2.018, && |m|a=0.0,\\
1.025&\lesssim\kappa a\lesssim2.180, && |m|a=0.2,\\
0.873&\lesssim\kappa a\lesssim2.346, && |m|a=1.0.
\end{aligned}
\label{eq:repulsive-intervals}
\end{align}
For these parameters, increasing $|m|a$ broadens the repulsive interval and shifts its onset toward smaller $|\kappa|a$. The intervals characterize scans of $\kappa$ at fixed $a$ and $m$. Table \ref{tab:casimir} illustrates why the force must be inferred from Eq. \eqref{eq:pressure-definition}. At $|m|a=0.2$ and $\kappa a=1$, the energy is positive while the pressure is attractive; at $|m|a=0$ and $\kappa a=2$, the energy is negative while the pressure is repulsive. The CFJ modification affects both the energy and its separation dependence. The resulting repulsion arises from the propagation and scattering of the coupled transverse modes, with its magnitude and parameter range controlled by the nonlocal response of the sheets.
\small{\begin{table}[!htbp]
\centering
\caption{Dimensionless energy and pressure obtained from the full Lifshitz determinant for the parameters of Fig.~\ref{fig:casimir}. Here $\mathcal{E}$ is the interaction energy per unit area.}
\label{tab:casimir}
\begin{tabular}{@{}|c|c|c|c|@{}}
\hline\hline
$|m|a$ & $\kappa a$ & $a^3\mathcal{E}$ & $a^4\mathcal{P}$\\ \hline\hline
%\midrule
$0.0$   & $0$ & $-6.65231\times10^{-5}$ & $-1.99569\times10^{-4}$\\
$0.0$   & $1$ & $-1.74582\times10^{-6}$ & $-2.16151\times10^{-5}$\\
$0.0$   & $2$ & $-2.63615\times10^{-6}$ & $+3.79362\times10^{-7}$\\
$0.2$ & $0$ & $-3.28586\times10^{-6}$ & $-1.34745\times10^{-5}$\\
$0.2$ & $1$ & $+1.93042\times10^{-7}$ & $-1.90308\times10^{-7}$\\
$0.2$ & $2$ & $-6.58622\times10^{-8}$ & $+3.85724\times10^{-7}$\\
$1.0$   & $0$ & $-3.89163\times10^{-7}$ & $-1.78751\times10^{-6}$\\
$1.0$   & $1$ & $+6.40296\times10^{-8}$ & $+1.76434\times10^{-7}$\\
$1.0$   & $2$ & $+1.34301\times10^{-9}$ & $+1.43817\times10^{-7}$\\
\hline
\end{tabular}
\end{table}}

\section{Conclusion}\label{sectionVII}

In this work, we investigated the quantum response and Casimir interaction of planar massive Dirac systems coupled to a Lorentz-violating vector sector described by a Bumblebee field supplemented by a CFJ term. The construction separates two physically distinct ingredients of the problem, i.e., the quantum response generated by the Dirac quasiparticles confined to the sheets and the modified propagation of the vector-field fluctuations in the bulk.

%Starting from the fermionic sector, we derived the complete one-loop polarization tensor associated with tangential Bumblebee fluctuations. For the space-like vacuum configuration normal to the sheets, the Bumblebee vacuum expectation value does not directly modify the planar quasiparticle spectrum, which remains that of a massive three-dimensional Dirac system. The polarization tensor was regularized and renormalized using the Pauli-Villars prescription while preserving the relations imposed by planar current conservation. The absorptive part displays the threshold at $\omega=2|m|$, while the gapless limit approaches the value $g^{2}/4$ in the Bumblebee channel. The parity-odd contributions cancel between the two valleys, and the dependence on the Fermi velocity disappears in the homogeneous optical limit. When the transverse Bumblebee mode is identified with the Maxwell-like photon, this response can be interpreted as the electromagnetic conductivity of the sheet.

The Casimir problem reveals a different role for the Lorentz-violating parameters. The bulk field equations and boundary conditions eliminate the normal fluctuation $\delta B_{3}$, so that the parameter $b$, associated with the spontaneous Lorentz-breaking vacuum, does not contribute to the interaction in the fluctuation sector. In contrast, the CFJ coefficient $\kappa$ remains dynamically relevant. It modifies the attenuation of the propagating modes and couples the TE and TM polarizations, producing nondiagonal propagation and reflection matrices. The corresponding real-frequency dispersion remains stable for the space-like background.

Using these bulk modes together with the full nonlocal response of the Dirac sheets, we obtained the Casimir energy and pressure within the Lifshitz scattering formalism while retaining the complete spatial dispersion and all orders of multiple reflection. In the Maxwell limit, $\kappa=0$, the Dirac gap progressively suppresses the interaction. At large separations, the opening of the gap changes the energy scaling from the gapless $a^{-3}$ behavior to an $m^{-2}a^{-5}$ decay, with the pressure correspondingly scaling as $m^{-2}a^{-6}$. This behavior follows from the suppression of the low-frequency surface response and shows how the Dirac mass directly modifies the long-distance vacuum interaction.

Meanwhile, the CFJ sector produces a qualitatively different effect. The Casimir energy is even under $\kappa\to -\kappa$, while the pressure develops a nonmonotonic dependence on $|\kappa|a$. For the parameter ranges analyzed, the attractive interaction is first weakened, then converted into a finite repulsive regime, and subsequently becomes attractive again. The Dirac mass also modifies these intervals. Increasing $|m|a$ broadens the repulsive region and shifts its onset toward smaller values of $|\kappa|a$. It is important to highlight that the sign of the interaction energy alone does not determine the sign of the force because the CFJ contribution modifies both the energy and its dependence on the separation. The repulsive behavior therefore originates from the combined propagation and scattering of the coupled transverse modes together with the nonlocal quantum response of the Dirac sheets.

These results show that the Casimir interaction in this system is governed by two complementary mechanisms. The Dirac gap controls the magnitude and long-distance scaling of the surface response, whereas the CFJ term modifies the bulk propagation and polarization mixing.

\section*{Acknowledgment}

This work was supported by the São Paulo Research Foundation (FAPESP) under Grant No. 2025/05176-7 (FCEL). The author also thanks Prof. D. Vassilevich for fruitful discussions.

\section*{Conflicts of interest}

The author declares no conflicts of interest.

\section*{Data availability}

No external datasets were used. The numerical data supporting the figures and table are generated from the equations presented in the manuscript.

\bibliographystyle{unsrt}
\bibliography{refs}

\end{document}